\documentclass[useAMS,usenatbib]{mn2e}

\input psfig.tex

\title[Galaxy evolution in clusters
up to $z=1.0$]{Galaxy evolution in clusters
up to $\bmath{z=1.0}$\thanks{Based on observations obtained at 
ESO and Cerro--Tololo telescopes.
}}

\author[S. Andreon et al.]
{S. Andreon,$^1$\thanks{email: andreon@brera.mi.astro.it}, 
J. Willis$^2$ \thanks{Present Address: Department of Physics
and Astronomy, University of Victoria, Victoria, Canada}, H. Quintana$^2$, 
I. Valtchanov$^3$\thanks{Present Address: Imperial College, London, UK}, 
M. Pierre$^3$ and F. Pacaud$^3$
\\
$^1$INAF--Osservatorio Astronomico di Brera, Milano, Italy \\
$^2$Departamento de Astronom\'\i a y Astrof\'\i sica, Pontificia Universidad Cat\'olica de Chile, Santiago, Chile\\
$^3$CEA/DSM/DAPNIA, Service d'Astrophysique, Gif-sur-Yvette, France\\
}
\date{Accepted ... Received ...}

\pagerange{\pageref{firstpage}--\pageref{lastpage}}
\pubyear{2003}

\begin{document}

\label{firstpage}

\maketitle

\begin{abstract}

We present a combined study of the colour--magnitude relation, colour
distribution and luminosity function of a sample of 24 clusters at
redshifts $0.3<z<1$. The sample is largely composed of X--ray
selected/detected clusters. Most of the clusters at redshifts $z<0.6$
display X--ray luminosity or richness typical of poor clusters or
groups, rather than the more typical, massive clusters studied in
literature at redshifts $z\ga0.3$. 
All our clusters, including groups, display a colour--magnitude relation
consistent with a passively evolving stellar population formed at a
redshift $z_f\ga2$, in accordance with observed galaxy populations in
more massive clusters studied at comparable redshifts.
Colours and luminosity functions show 
that the cluster galaxy  population is consistent with the
presence of at least two components: old systems formed at
high redshift that have evolved passively from that epoch, together
with a galaxy population displaying more recent star formation.
The former population forms at $2\la z_f \la 5$, the latter at redshifts $z<1$.
A model in which stars do not evolve is clearly rejected by both by the
colour of reddest galaxies and by the characteristic luminosity $m^*$ 
measures.
All clusters (with one possible exception) are detected independently
by an almost three dimensional optical search employing sky position
and colour -- this despite the primary X--ray selection and low
X--ray flux/optical richness displayed by the majority of the sample.

\end{abstract}


\begin{table*}
\caption{Summary of the cluster sample considered in this paper.}
\begin{tabular}{llcclll}
\hline
 Name & Colour  &\multispan{2}{\hfill Optical centre (J2000) \hfill} & ~~z & Filters & Notes\\
      & detected? & RA & DEC  & & & \\ 
\hline
XLSSC 008  & -- & 02~25~20.2 &  -03~48~30 & 0.297  &   $I$     &   \\
XLSSC 013  & Y  & 02~27~25.9 &  -04~32~15 & 0.307  &   $Rz'$	&   \\
XLSSC 018  & Y  & 02~24~01.6 &  -05~05~25 & 0.322  &   $Rz'$	&   \\
XLSSC 009  & -- & 02~26~44.7 &  -03~41~02 & 0.327  &   $I$     &   \\
XLSSC 010  & -- & 02~27~22.3 &  -03~21~41 & 0.329  &   $I$     &   \\
XLSSC 016  & Y  & 02~28~28.2 &  -04~59~46 & 0.332  &   $Rz'$	&   \\
XLSSC 014  & Y  & 02~26~34.5 &  -04~03~55 & 0.344  &   $Rz'$   &   \\
XLSSC 017  & Y  & 02~26~27.4 &  -04~59~55 & 0.381  &   $Rz'$   &  blended with XLSSC 020 \\
VMF98 34   & Y  & 03~41~57.0 &  -45~00~11 & 0.408  &   $Rz'$   &  X--ray selected \\
XLSSC 006  & Y  & 02~21~45.8 &  -03~46~08 & 0.429  &   $RIz'$  &   \\
XLSSC 012  & Y  & 02~28~27.4 &  -04~25~48 & 0.433  &   $Rz'$   &   \\
VMF98 43   & Y  & 05~29~38.0 &  -58~48~20 & 0.466  &   $Rz'$   &  X--ray selected \\
RzCS  001  & Y  & 02~24~04.3 &  -05~17~22 & 0.494  &   $RIz'$  &  colour selected, X--ray undetected \\
XLSSC 019  & Y  & 02~24~11.8 &  -05~22~47 & 0.494  &   $Rz'$   &   \\
XLSSC 020  & Y  & 02~26~32.8 &  -05~00~32 & 0.494  &   $Rz'$   &  blended with XLSSC 017 \\
Cl0412     & Y  & 04~12~49.9 &  -65~50~44 & 0.51   &   $Rz'$   &  optically selected, alias F1557.19TC \\
XLSSC 007  & N	& 02~24~09.0 &  -03~55~09 & 0.557  &	$RIz'$  & dubious X--ray-optical identification\\
VMF98 40   & Y  & 05~21~12.0 &  -25~31~13 & 0.581  &   $Rz'$   &  X--ray selected \\
XLSSC 001  & -- & 02~24~57.1 &  -03~48~53 & 0.614  &   $I$     &   \\
XLSSC 002  & -- & 02~25~32.5 &  -03~55~10 & 0.772  &   $I$     &   \\
RXJ0152    & Y	& 01~52~43.9 &  -13~57~19 & 0.831  &   $Rz'$	&  X--ray selected, alias RX J0152.7-1357\\
XLSSC 003  & -- & 02~27~37.6 &  -03~18~07 & 0.838  &   $I$     &   \\
XLSSC 004  & Y  & 02~25~28.4 &  -05~06~57 & 0.88   &   $RIz'$  &   \\
XLSSC 005  & Y  & 02~27~09.7 &  -04~18~05 & 1.0    &   $RIz'$  &   high z structure\\
\hline							 
\end{tabular}
\medskip
\break \noindent
Notes: redshift for XLSSC clusters are taken from Valtchanov et al. (2003) \& 
Willis et al. (2004). VMF98 34, VMF98 40 and VMF98 43, 
are drawn from the 160 deg$^2$ survey (Vikhlinin et al.
1998, Mullis et al. 2002), while RXJ0152 is drawn from the
SHARC survey (Romer et al. 2000), and also detected in the WARPS
survey (Ebeling et al. 2000). Cl0412 is an optically
selected cluster (Couch et al. 1991) later observed (and detected) in X--ray
(Smail et al. 1997). \hfill \quad				 
\end{table*}

\section{Introduction}

The colour--magnitude relation and the luminosity function (LF)
provide quantitative measures of galaxy evolution.  The
colour--magnitude relation, also known as the red sequence, is a
general observed characteristic of galaxies in clusters
(e.g. Ga\-ril\-li et al. 1996, Stanford, Eisenhardt \& Dickinson
1998). The homogeneity in colour of galaxies on the red sequence, both
observed across a range of clusters (Ellis et al. 1997; Andreon
2003a,c) and within individual systems (e.g. Bower et al. 1992,
Stanford et al. 1998), and the apparent passive evolution  
of cluster elliptical galaxies 
(Kodama \& Arimoto 1997, Stanford et al. 1998; Kodama et
al. 1998), all imply that the luminosity--weighted stellar populations
within such galaxies are uniformly old ($z_f\ga2$).

The colour--magnitude relation constrains the evolution of the reddest
cluster galaxies whereas the LF describes the spatial density per unit
luminosity interval and its evolution provides an overall measure of
the changing cluster galaxy population.  The local ($z<0.3$) cluster
galaxy LF has been derived for large numbers of systems (Garilli,
Maccagni \& Andreon 1999; Paolillo et al. 2001; de Propris et
al. 2003) and the evolution of the LF at increasing redshift (Andreon
2004) is consistent with predictions based upon passive stellar
evolution. Near infra--red (NIR; in this case $K$--band) observations
have confirmed that the mean cluster galaxy LF continues to evolve
passively to redshift unity (de Propris et al. 1999). NIR fluxes
received from galaxies at redshifts $z<1$ are dominated by rest--frame
emission arising from stellar types G and later. Therefore, although
the NIR cluster galaxy LF provides a suitable measure of the passive
evolution of stellar mass contained in such systems, the LF computed
from red optical passbands (i.e. sampling rest--frame blue emission)
provides a more sensitive measure of active luminosity evolution,
i.e. secondary star formation events.

Current measures of the optical LF evolution of cluster galaxies at
$z>0.3$ are few (Nelson et al. 2001; Barrientos \& Lilly 2003).  
The present paper addresses this
issue and combines a discussion of LF evolution in cluster galaxies
with a simultaneous assessment of the colour--magnitude relation.
Data acquisition and reduction is discussed in Section 2.  Section 3
presents the colour--magnitude relation, colour distribution and the
LF computed for individual clusters. Section 3 also discusses the
evolution of the above quantities with redshift.  The main results are
presented and discussed in Section 4, i.e.  we demonstrate that the
cluster galaxy population is consistent with the presence of at least
two components: old systems formed at high redshift that
have evolved passively from that epoch, together with a galaxy
population displaying more recent star formation. 
We discuss constraints placed on the
evolution of both populations by the current data set.

Throughout this paper we assume a
Friedmann--Robertson--Walker--Lemaitre cosmological model described by
the parameters $\Omega_M=0.3$, $\Omega_\Lambda=0.7$ and $H_0=70$
kms$^{-1}$ Mpc$^{-1}$.

\section{The cluster sample: Observations, data reduction and colour analysis}

\subsection{The cluster sample}

The cluster sample presented in this paper is drawn from a number of
sources and represents a heterogeneous data set. Uniform photometry
was obtained for all clusters and details are presented in Table
1. Eighteen clusters were observed as part of the X--ray Multi--Mirror
(XMM) Large Scale Structure (LSS) survey (Pierre et al. 2004) which
aims to determine the large scale structure of the Universe as traced
by galaxy clusters\footnote{The XMM--LSS survey area is centred on
the coordinates $\alpha=02:18:00$, $\delta=-07:00:00$ (J2000).}.  An
additional 5 clusters were added to this sample from the
literature. The majority of the clusters are drawn from X--ray
selected samples; two (RzCS 001 \& Cl0412) are optically
selected, the latter being later X--ray detected.  
The nature of the most distant cluster, XLSSC 005, is
ambiguous (Valtchanov et al. 2004). 
Although the system displays extended X--ray emission and a
concentration of galaxies in redshift space, there are several other
galaxies in the cluster surroundings ($0.92<z<1.05$), suggesting
perhaps a more complex structure. The redshift thickness of this
structure is negligible, and therefore neglected within our subsequent
analyses.

\begin{figure}
\psfig{figure=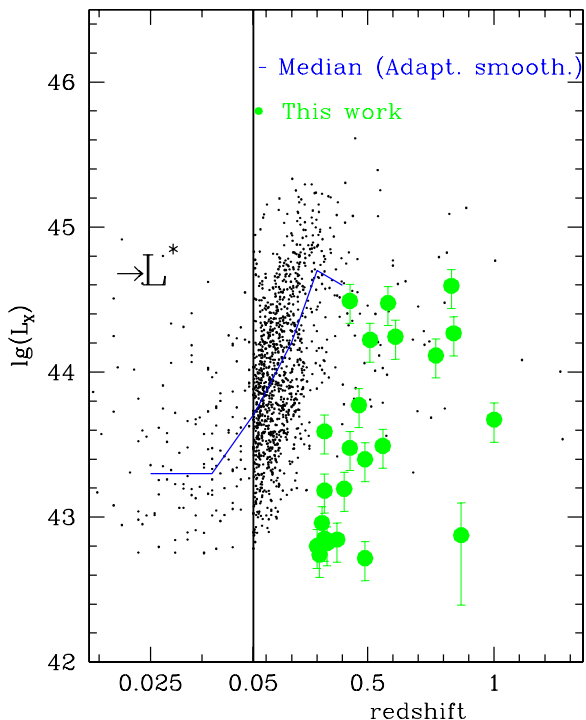,width=8truecm,clip=}
\caption[h]{X-ray luminosity measured in the [0.1-2.4] keV rest--frame
versus redshift for clusters drawn from literature (black dots) and
XLSSC clusters (green circles with error bars). The blue curve is an
adaptively smoothed running median of literature points.}
\end{figure}

Figure 1 shows the X--ray luminosity in the [0.1-2.4] keV rest--frame
band versus redshift for more than 1000 clusters in literature and
listed in the X-Ray Galaxy Clusters Database (BAX, black dots), 
and for all X--ray detected clusters
studied here (green circles with error bars). 
The horizontal arrow indicates the
characteristic $L^*_X$ luminosity reported by Ebeling et
al. (1997). The curve is an adaptively smoothed running median of
literature points.  The X--ray luminosity of the clusters considered
in this paper are typically lower than clusters presented in the
literature (e.g., Mullis et al. 2003; Romer et al. 2000; Ellis et
al. 1997) at similar redshift.  In particular, at redshifts $z<0.6$,
XLSSC clusters have X--ray luminosities characteristic of low mass
clusters and groups (Willis et al. 2004). This is expected, given the
present area coverage and limiting flux of the XMM--LSS project
(Pierre et al. 2004).

All clusters presented in this paper (including the complex structure
XLSSC 005) have at least two concordant redshifts in addition to the
presence of an unambiguous galaxy overdensity in multi--colour CCD
images. We note that, though cluster XLSSC 007 is confirmed
spectroscopically (Willis et al. 2004) with at least 10 members, there
exists a large (1.2 arcminute) offset between the optical galaxy
overdensity and the centroid of the X--ray emission.  Though the
optical cluster is clearly real, the X--ray flux for XLSSC 007 will be
greatly overestimated should the observed X--ray emission arise from a
second cluster along the line of sight.

\subsection{Observations and data reduction}

All clusters presented in this paper have been observed in $R,z$
and/or $I$ passbands as detailed in Table 1. Optical $R$-- and
$z'$--band ($\lambda_c\sim9000${\AA}) images were obtained at the
Cerro Tololo Inter--American Observatory (CTIO) 4m Blanco telescope
during two observing runs, in August 2000 and November 2001, with the
Mosaic II camera. Mosaic II is a 8k$\times$8k camera with a $36 \times
36$ arcminute field of view. Typical exposure times were 1200 seconds
in $R$ and $2 \times 750$ seconds in $z'$. Seeing in the final images
was between 1.0 and 1.4 arcseconds Full--Width at Half--Maximum (FWHM)
in the November 2001 run (when all clusters except XLSSC 006 were
observed), and 0.9 to 1.0 arcsec FWHM in the August 2000 run (XLSSC
006 observations). The useful nights of the two observing runs were
photometric. Images were trimmed and bias corrected. A flat field
correction was applied together with an illumination correction (where
required) and interference fringes were removed from $z'$--band
images. In the data reduction we employ the {\tt FLIPS} software
package (Cuillandre, in preparation). Where multiple exposures of the
same field were available, cosmic rays were identified and images were
combined. Extensive comparisons to Landolt (1992) standard stars, sky
regions in the Early Data Release of the Sloan Digital Sky Survey
(SDSS; Stoughton et al. 2002), and overlapping regions between
pointings demonstrated that the photometric zero point is accurate to
better 0.03 mag over the entire instrument field of view. A detailed
description of these data reduction techniques will presented in a
forthcoming paper presenting optically selected galaxy clusters within
the XMM--LSS survey (Andreon et al. 2004).

Additional $I$--band images were obtained at the European Southern
Observatory (ESO) 8.2m Very Large Telescope (VLT) facility employing
the FOcal Reduction Spectrograph (FORS2) in September 2002 as part of
pre--imaging of spectroscopic target fields. The FORS2 instrument
consists of two 2k$\times$4k CCDs with a field of view of 7$\times$7
arcminutes. Exposure times were either $2 \times 150$ seconds or $4
\times 150$ seconds and observations were performed under photometric
sky conditions. FORS2 images were reduced using standard techniques
using IRAF\footnote{IRAF is distributed by the National Optical
Astronomy Observatories, which are operated by the Association of
Universities for Research in Astronomy, Inc., under cooperative
agreement with the National Science Foundation.}. Seeing in the final
images were $0.6-0.9$ arcseconds FWHM.

Object magnitudes are quoted in the photometric system of the
associated standard stars: $R$ and $I$ magnitudes are calibrated with
Landolt (1992) stars, while $z'$ magnitudes are calibrated with SDSS
(Smith et al. 2002) standard stars.  Source detection and
characterization was performed employing SExtractor v2 (Bertin \&
Arnouts 1996). Colours are computed within a fixed 1.9
arcsecond radius aperture, whereas magnitudes are computed
within an angular aperture of projected size equal to 15.3 kpc radius
for objects at the cluster redshift.

A fixed angular aperture is employed to compute object colours,
irrespective of the object redshift. We adopt this approach as only
limited spectroscopic information is available within each cluster
field.  Employing a fixed angular aperture translates to a uniform
metric aperture for all galaxies located at the cluster redshift. 
Colours computed within the adopted 1.9 arcsecond aperture are biased
by differential seeing effects between the $R$-- and $z'$--band
images.  A bias correction is applied employing bright stars located
within each field.  Several possible angular radii were considered,
and 1.9 arcseconds was selected as a compromise figure that generated
acceptable source signal--to--noise ratios together with a small
correction for differential seeing effects. The median absolute
correction (over all CTIO pointings) 
is 0.05 mag and the scatter is 0.05 mag.

A fixed metric aperture, corresponding to a radius of 15.3 kpc at each
cluster redshift is employed to determine galaxy photometry for the LF
computation. The particular aperture applicable to each cluster is
also applied to galaxies within the control field.
This metric aperture is employed in order to avoid introducing an
unnecessary bias associated with the use of non--metric apertures
(Dalcanton 1998) which can mimic the effect of redshift evolution in
the LF.  The value for the aperture radius, 15.3 kpc, permits a consistent
comparison between the results of the current study and the LF derived
for 65 low redshift ($z<0.25$) clusters by Garilli, Maccagni \&
Andreon (1999 -- hereafter GMA99), that adopt the same aperture when
cast within their adopted cosmological model.

\begin{figure*}
\psfig{figure=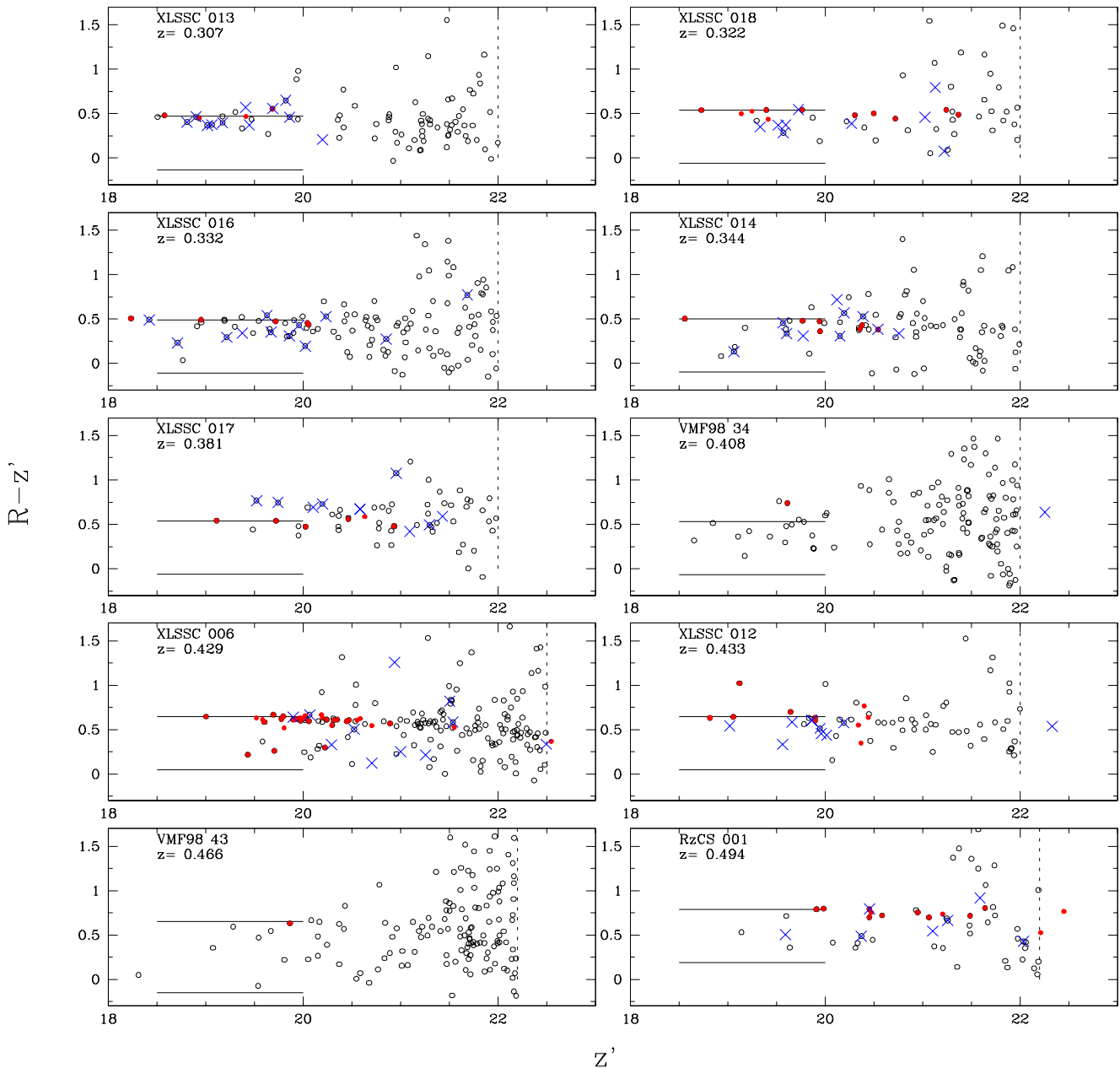,width=18truecm,clip=}
\caption[h]{Colour--magnitude diagram for galaxies (open circles)
brighter than the $z'$ magnitude limit (vertical line). The horizontal
lines indicate the colour range adopted for the LF derivation. Most of
the spectroscopically confirmed members (red filled circles) are
located within this colour range. Interlopers (blue crosses) are often
located outside this colour range (e.g. XLSSC 004 or XLSSC 020). A
small number of cluster members are scattered to colours significantly
redder than this colour range, mainly as a result of source crowding
and consequent deblending problems.}
\end{figure*}

\begin{figure*}
\contcaption{}
\psfig{figure=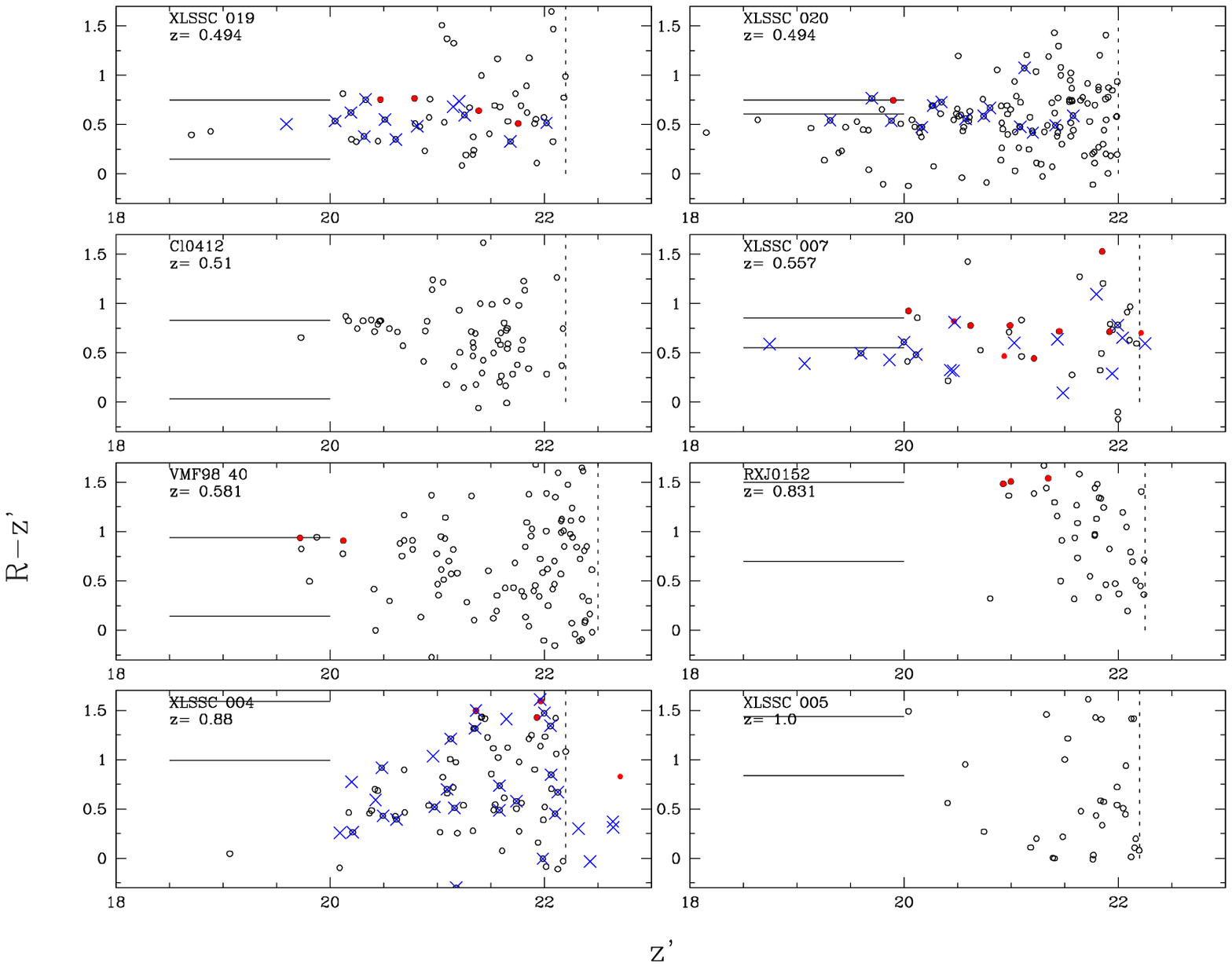,width=18truecm,clip=}
\end{figure*}

At faint magnitude limits, galaxies displaying low central surface
brightness values are detected with rapidly decreasing
frequency. Aperture magnitude completeness limits have been estimated
according to the prescription of Garilli, Maccagni \& Andreon (1999),
Andreon et al. (2000), and Andreon \& Cuillandre (2002), by
considering the magnitude of the brightest galaxies displaying the
lowest detected central surface brightness values. Only galaxies
brighter than this completeness limit are considered in this paper.

Bright objects (typically $R<22$, $z'<22$), whose compactness computed
using the SExtractor stellar classifier leads to a relatively
unambiguous stellar classification, are discarded from the galaxy
sample. Fainter stars are subtracted in a statistical manner following
Andreon \& Cuillandre (2002). The statistical nature of this approach
avoids incorrectly excluding faint, compact galaxies from the
sample. The control field used in the statistical subtraction
procedure is separated from the science fields by no more than one
degree and, as observations are performed at high Galactic latitudes,
star counts within the control field are expected to reproduce those
of the science fields.

\subsection{Colour analysis and cluster detection}

Galaxy clusters have been detected employing a method similar to the
red sequence method of Gladders \& Yee (2000). The method employed in
this paper differs from the Gladders \& Yee (2000) approach in several
key areas, as described when it was applied (Andreon 2003a,b) to the
SDSS Early Data Release (Stoughton et al. 2002).  In summary, the
method exploits the observed trend that the majority of galaxies in
clusters display similar colours, while non--cluster
galaxies located along the line--of--sight display considerable
variation of observed colours, both because they are drawn from a
larger interval of redshift and because the field galaxy population at
a given redshift displays a larger variation in colour than a typical
cluster galaxy population. The algorithm identifies local galaxy
overdensities displaying similar $R-z'$ colours, and is considered at
several angular scales.  The applied colour filtering effectively
removes most of the ``background'' galaxies
(see Figure 3 and later discussion).

Only one cluster (XLSSC 007), of the 25 clusters studied here, is
undetected at the $1-5 \times 10^{-6}$ confidence level. The remaining
clusters present such sufficiently well--characterised detections that
the optical images used to detect each cluster also permit detailed
colour and LF studies.

\section{Results}

\subsection{The colour--magnitude distribution}

Figure 2 displays colour magnitude diagrams constructed for clusters
listed in Table 1 with $R$-- and $z'$--band photometry.  Two sample
distributions are displayed: The first sample (open points) consists
of galaxies brighter than the completeness limit (vertical dashed
line) located within a specified radius from the cluster centre.  This
radius is specified to optimize the signal to noise ratio of the
resulting LF computation. A radius of 2 arcminutes was adopted for the
majority of the clusters. Exceptions to this value include clusters
with a large projected extent (VMF98 34, VMF98 43, XLSSC 004, XLSSC
014, XLSSC 016, and XLSSC 020), for which a 3 arcminute radius was
applied.  In addition, a 1.5 arcminute radius was applied in two
cases: XLSSC 017 -- to reduce any contamination from the nearby
cluster XLSSC 020; and XLSSC 007 -- to reduce any effect from an
unassociated foreground galaxy located 2 arcminutes from the cluster
centre.  The same set of apertures are employed to compute both the
cluster colour distributions and the LFs.  An angle of two arcminutes
corresponds to a distance of $0.53,0.80,0.96$ Mpc when projected
at redshifts $z=0.3,0.6,1$ respectively.

The red sequence within each cluster is clearly apparent at redshifts
$z<0.6$, although a comparison of individual clusters demonstrates
that their visual appearance varies considerably (e.g. XLSSC 006
compared to VMF98 43). The red sequence is sampled over an
increasingly limited magnitude range with increasing redshift and is
more heavely contaminated by background galaxies. However, the sequence
remains clearly identifiable to redshifts $z=0.84$.

The red envelope of the red sequence is defined by the reddest
(i.e. larger colour value) horizontal line displayed in each
sub--panel of Figure 2. The value of the red envelope is derived from
the median colour of the three brightest galaxies considered to be
viable cluster members, i.e. galaxies that are too blue or too bright
to be plausibly at the cluster redshift are discarded.

\begin{figure*}
\psfig{figure=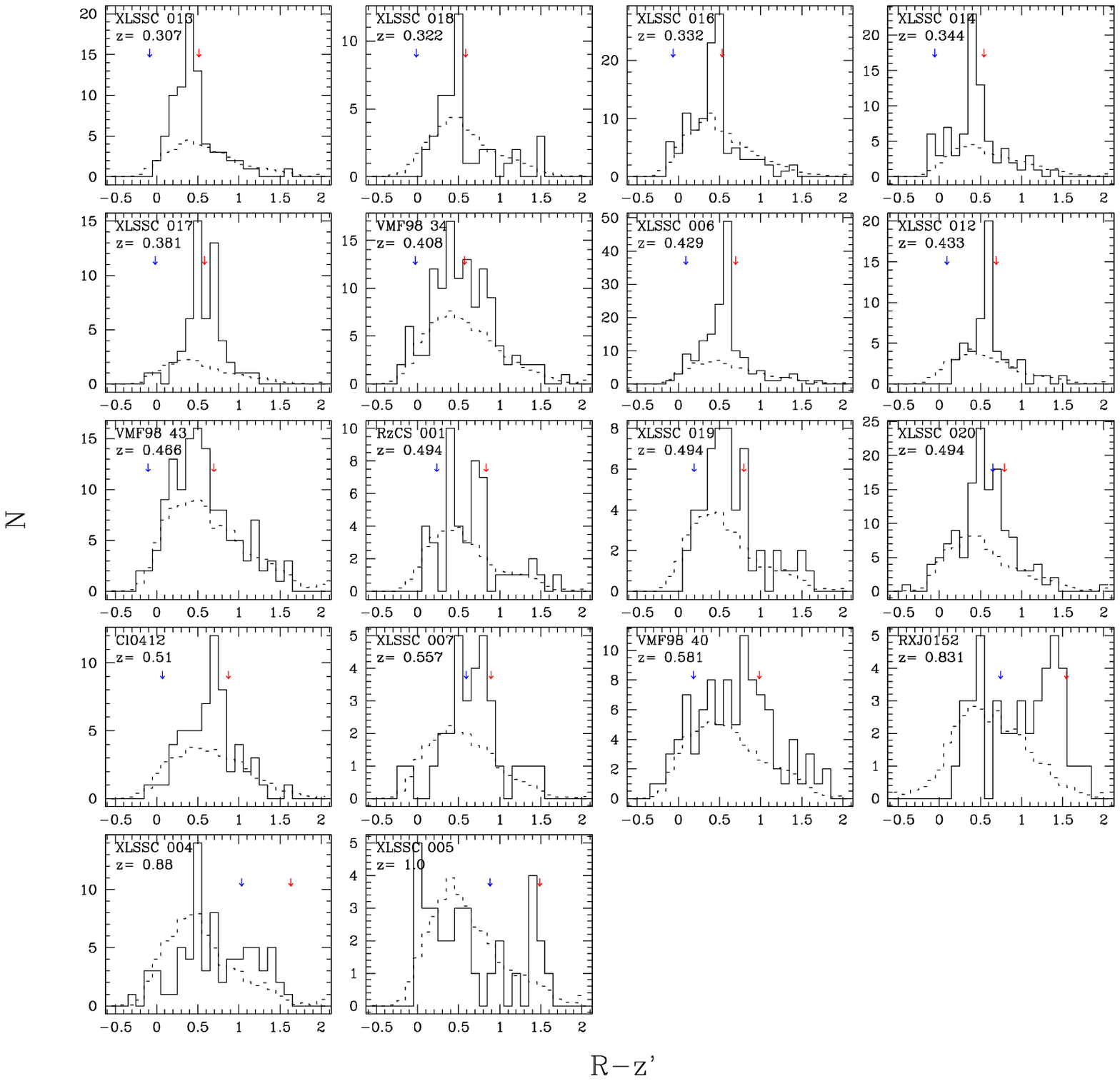,width=18truecm}
\caption[h]{Colour histograms of galaxies brighter than the $z'$
completeness magnitude limit in each cluster field. The solid
histogram refers to the cluster field, while the dashed histogram
indicates the colour distribution of the control field normalized to
the cluster field area. The arrows mark the colour range adopted for
the LF derivation. There is no statistically significant (at the
99.9995 \% or greater confidence level) excess of galaxies outside this
range, except in obvious cases (e.g. when another cluster is located
along the line of sight).}
\end{figure*} 

The second sample of objects displayed in Figure 2 consists of a
heterogeneous sample of galaxies with spectroscopic redshifts
(Valtchanov et al. 2004; Willis et al. 2004; supplemented by NED and
private communications for clusters drawn from literature), selected
with no constraint with respect to $z'$ magnitude or distance to the
defined cluster centre. A galaxy is defined as a cluster member if
lies within three $\sigma_v$ of the centre of the cluster
line--of--sight velocity distribution.  Exceptions to this criterion
are clusters lacking a well--defined $\sigma_v$ value (literature
clusters and XLSSC 004), for which a value of $\sigma_v=1000$
kms$^{-1}$ is adopted. Galaxies with spectroscopic redshifts that do
not satisfy this criterion are considered interlopers. Due to the
complex nature of the system XLSSC 005, we do not include objects with
spectroscopic redshifts in the discussion of this system. 
The distribution of spectroscopic cluster members (red points)
coincides with and reinforces the colour--magnitude relation defined
by the photometry alone.

\subsection{Interpreting the colour distribution for each cluster}
\label{subsec_col_dist}

The colour distribution of galaxies brighter than the $z'$--band
completeness limit within each cluster field is displayed in Figure 3.
In each case, the colour distribution within the cluster radius
defined in the previous section is compared to the colour distribution,
normalized to the cluster area, of the whole MOSAIC II image
(36 $\times$ 36 square arcminutes) in which the cluster is observed
At redshifts
$z\la0.8$, a significant excess of galaxies is observed toward the
cluster field compared to the control field -- whether computed over a
specific colour interval or integrated over the entire colour
distribution.  The excess due to the cluster is sufficiently
significant that clusters should be apparent as a spatial overdensity
in a $z'$--band catalogue alone.  At redshifts $z\ga0.8$, the excess
of galaxies in colour space due to the cluster compared to the
background is only significant within a limited (red) colour interval.
These clusters are unlikely to be identified via a galaxy overdensity
in a single photometric band, i.e. by neglecting the colour
information.  All clusters within the sample display a significant
numerical excess over a limited colour interval (typically $\pm 0.3$
mag), indicating that clusters may be identified effectively at
redshifts $z\ga0.8$ by methods that employ colour selection to
suppress background galaxy signals (Section 2.3).  All clusters, with
the exception of XLSSC 007, were in fact detected in a 3 dimensional
space defined by sky position and $R-z'$ colour (Table 1).

A local ($z<0.34$) cluster comparison sample, detected using a
preliminary version of the same cluster detection algorithm used in
this paper, is shown in Figure 1 of Andreon (2003a). The colour
distribution of this local sample is qualitatively quite similar to
the one shown in Figure 3. A detailed study of the evolution of the
colour distribution (i.e. on the Butcher--Oemler effect) will be
presented in a later study.

The cluster sample presented in this paper is predominantly X--ray
selected/detected.  The identification of a red sequence of galaxies
associated with each cluster would initially seem at variance with
Donahue et al. (2002), who claim that X--ray clusters presented in
their survey do not all display a prominent red sequence.  However, as
the latter authors note and Gladders \& Yee (2000) show, clusters do
not require a prominent red sequence to be detected by a methods that
search for overdensities of galaxies of similar (but not identical)
colour. This refinement of the definition of the red sequence would
appear to resolve the apparent contradiction between our finding and
Donahue et al. (2002) claim and confirming the recent results by
Gilbank, Bower, Castander, \& Ziegler (2003), based on a low ($z<0.4$)
redshift sample.

\begin{figure*}
\psfig{figure=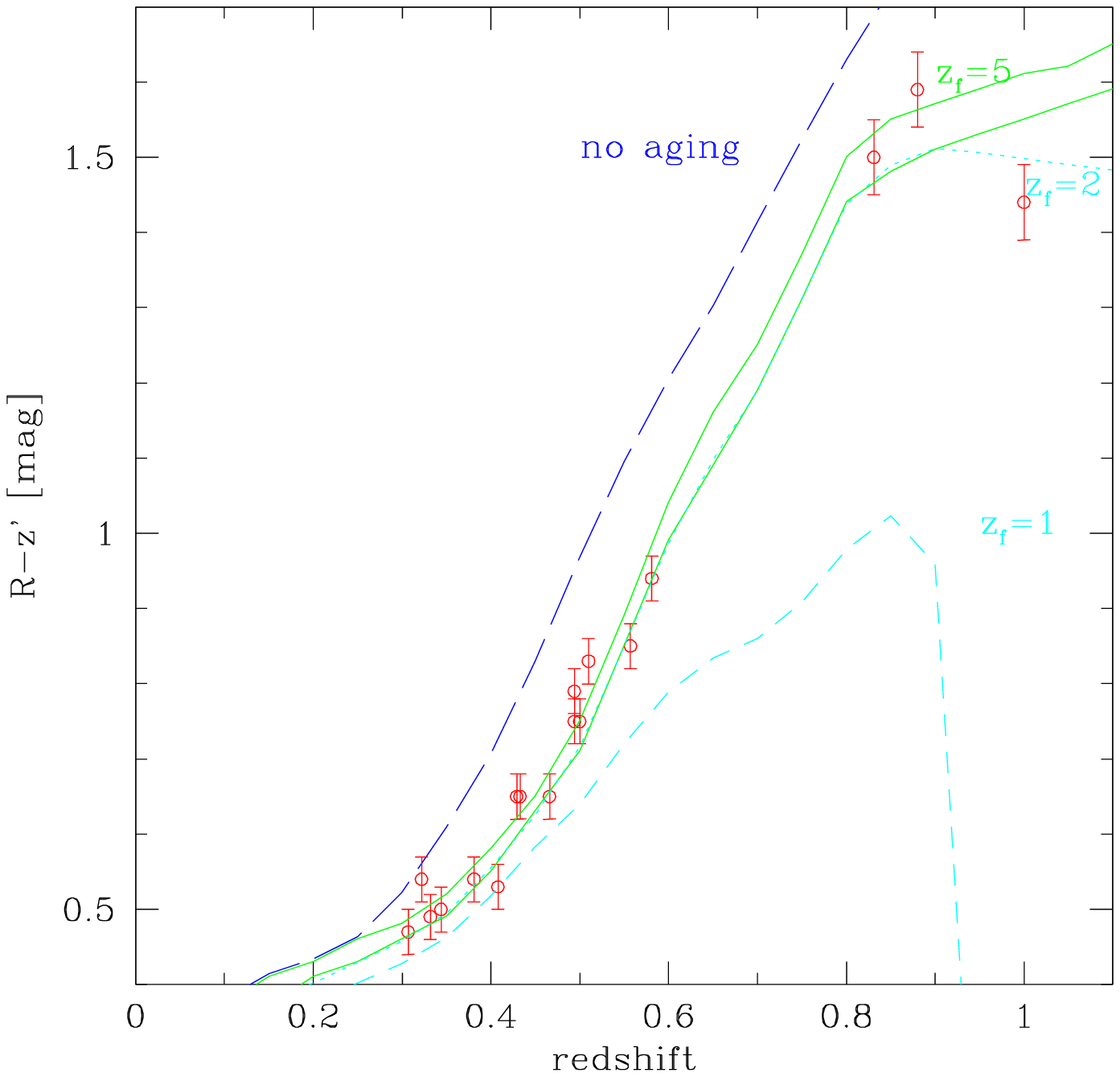,width=12truecm}
\caption[l]{Observed $R-z'$ colour of the red envelope of the red
sequence observed in each cluster as a function of redshift. Three
galaxy evolution models are considered: a non--evolving early--type
galaxy of present--day age at all redshifts, and two evolving
early--type galaxy models, each characterised by a different formation
redshift and mass. See text for details.}
\end{figure*}

The clusters presented in this paper display a range of masses (as
determined by either dynamical or X--ray information, or both).  In
particular, XLSSC clusters at $z<0.6$ display X--ray luminosities
comparable to low richness clusters or groups (Figure 1).  Therefore, the
detection of galaxy overdensities in the 3--space defined by colour
and sky location, at the location of extended X--ray sources indicates
that such techniques may provide a promising route to confirm the
nature of low mass X--ray selected clusters.

However, colour plus position selection alone does not constrain the
extension in redshift of the identified structure. A filamentary
structure of galaxies seen along the line of sight is, without
spectroscopic data, hard to distinguish from a cluster and both
scenarios can in principle give rise to the ``cluster"
detection. Spectroscopic observations of a sample of colour selected
structures are therefore required to measure the frequency of each
type of structure (clusters versus non--virialised large--scale
structure).  The spectroscopic cluster sample presented in this paper
contains a colour selected cluster undetected in X--ray
(RzCS 011). This system is confirmed spectroscopically
and displays a well--defined mean redshift and distribution of
rest--frame velocities. It is, therefore, a cluster in the sense of
being a gravitationally bound systems of galaxies, altough 
undetected in X--rays.  Therefore, colour
selection techniques provide a method to identify clusters displaying
a broad range of X--ray properties, possibly
sampling the cluster mass function deeper than X--ray observations. 
A more complete investigation of
the bivariate distribution of optical and X--ray properties of distant
clusters await a larger sample.

\subsection{Colour evolution of the red sequence}

The evolution of the color of the red sequence, as
computed from the median colour of the three brightest galaxies on
the red sequence (Figure 2) is displayed in Figure 4\footnote{The data
points are available in electronic form
at the URL http://www.brera.mi.astro.it/$\sim$andreon/XIDindex.html}. The colour error on each data point is estimated to be 0.03 mag,
based upon the present estimate of the maximal variation of the
photometric (flux) calibration across the camera field of view
(Section 2). The three highest redshift clusters are assumed to
exhibit a colour error 0.05 mag, due to the lower signal--to--noise
ratio of the photometry.

The colour of the red envelope becomes monotonically redder with
increasing redshift (to at least redshifts $z\sim1$). Several model
predictions are indicated in Figure 4: the top (long dashed) curve
neglects aging of the stellar population. It is computed assuming a
non--evolving 12 Gyr old early--type galaxy SED produced using the
GISSEL98 spectral library (Bruzual \& Charlot 1993). Galaxy colours
are computed by convolving the spectrum with the appropriate filter
transmission function together with the atmospheric transmission
spectrum. Colours are zero--pointed to match the colour of early--type
galaxies within the Coma cluster -- the models of Kodama \& Arimoto
(1997) are employed for this purpose to avoid uncertainties due to the
differences between the filters employed in this paper and those used
for Coma galaxies by Bower, Lucey \& Ellis (1992).
 
The data points are clearly inconsistent with a universe where the
oldest stars have the same age at all redshifts, as already shown by
Kodama et al. (1998).  The additional models indicated in Figure 4 are
more physically motivated.  Passive stellar ageing and chemical
evolution are incorporated employing the model of Kodama \& Arimoto
(1997), that assumes a formation redshift, $z_f$, and a total stellar
mass. The two solid green curves indicate $z_f=5$ and a total
stellar mass of $\sim 1.7 \times 10^{11} M_\odot$ and $\sim 6.4 \times
10^{10} M_\odot$.  The expectation for a mass of $\sim 6.4 \times
10^{10} M_\odot$ and two lower formation redshifts ($z_f=2$ and
$z_f=1$) are plotted as dotted and short dashed curves respectively.

The colour of the envelope of the red sequence is reproduced well by
models where the oldest stars form at $2\la z \la5$, in good agreement
with the findings of Stanford et al. (1998) and Kodama et al. (1998)
based on a set of richer clusters located within a comparable redshift
range, and with Andreon et al. (2003a,b) for a sample of low redshift
clusters of low optical richness, and also with Aragon--Salamanca et
al. (1993) for a near--infrared study.

No clusters within the sample, particularly in the redshift range
$0.3<z<0.6$ where the colour--magnitude relation is well--sampled by
the observations, displays a significant deviation
(i.e. $\Delta(R-z')>0.1$) from the average trend, i.e. no cluster
displays an unusually red or blue colour--magnitude relation for its
redshift. This observation is similar to that noted at $z<0.34$ in a
sample of more than 150 clusters (Andreon 2003a,b).

The high formation redshift for red sequence galaxies (alternatively,
the old age of the constituent stellar populations), although similar
to previous studies, displays a number of important new aspects.  Firstly,
the observations presented form a very homogeneous data set: all
clusters (with the exception of XLSSC 006) were observed with the same
instrument and filters during a single observing run (XLSSC 006 was
observed using the same instrument on the previous run).  By way of
comparison, the cluster samples presented by Stanford et al. (1998)
and Kodama et al. (1998) employ a combination of telescopes and
photometric passbands. Secondly, the cluster sample is 
mostly X--ray selected. The cluster sample is therefore selected
in a manner which is independent of (or assumed to be) optical
properties and avoids a potentially circular analysis between
optically selected clusters and the properties of the red
sequence. Thirdly, many of the clusters at $0.3<z<0.6$
presented within this paper display X--ray luminosities (and richness,
see section 3.4) indicative of low mass or low richness clusters and
groups rather than optically rich, massive clusters considered in
previous studies (though c.f. Andreon 2003a,b). Low mass structures
represent environments with current predictions regarding the assembly
of bright, red galaxies (e.g. Kauffmann 1996 and Eggen, Lynden-Bell \&
Sandage 1962) are deemed to display the greatest divergence.

We note that the formation redshift computed for the sample depends in
a non--trivial manner upon the assumed stellar population model:
adopting a Bruzual \& Charlot (1993, updated to GISSEL96) model of
solar metallicity, a Salpeter IMF and a star formation e--folding time
scale $\tau=1$ Gyr, causes the formation redshift that reproduces the
colour of the red sequence of the cluster sample to shift from
redshifts $2 \la z_f \la 5$ to $z_f\sim11$. Stanford et al. (1998)
employed GISSEL96 and reported formation redshifts for early--type
galaxies located on the red sequence of $2 \la z_f \la 5$.  It is
important to note that differences in the cosmological model assumed
in this paper and that adopted by both Kodama et al. (1998) and
Stanford et al. (1998) mean that the formation redshifts computed by
these authors should be lowered somewhat when compared to the
current results.

The presence of colour gradients within cluster galaxies is not
expected to alter the conclusions regarding the colour evolution of
the red cluster galaxy population.  This assumption holds even in the
current situation where the fixed angular aperture employed to measure
cluster galaxy colours corresponds to a varying rest--frame aperture
projected at the redshift of each cluster.  There are several reasons
for this expectation: 1) differential colour gradients, $\partial
(B-R) / \partial \log r$, are small (e.g. 0.02 mag per decade in
radius, Vader et al. 1988), and the resulting integrated colour
gradient $\partial (B-R)(<r) / \partial \log r$ displays an even
smaller dependence upon radius, 2) the applied angular aperture
changes by a factor 2 (1/5 of a decade) in the galaxy rest--frame as
the galaxy is moved from $z=0.3$ to $z=1.0$ and 3) the aperture is
intrinsically large compared to the visible galaxy extent at all
redshifts considered in the sample (about 30 kpc once seeing effects
are considered -- see also Kodama et al. 1988).

\subsection{Cluster luminosity functions}

The cluster LF is computed employing standard techniques (e.g. Oemler
1974), i.e. galaxy counts are compiled from the cluster field and a
background galaxy count computed from a control field, suitably
normalized to the cluster area, is subtracted.  The effective cluster
area is further corrected for the crowding effect due to bright stars
where required (e.g. for Cl0412). Errors on the LF data points take
into account the increased variance in galaxy counts due to large
scale structure variations and follow the Huang et al. (1997)
approach. We note that several literature papers do not take such
sources of variance into account, and assume Poissonian uncertainties
alone. Taking such additional sources of uncertainty into account can
give rise to the misleading impression of a lower data quality when
compared to quoted uncertainties for literature LFs. For example,
Barrientos \& Lilly (2003) do not include a term associated with
large--scale structure variation in the LF uncertainty. As emphasized
by Andreon \& Cuillandre (2002), background fluctuations enter into
the error budget twice -- one contribution from the control field and
one from the cluster direction field.

A single Mosaic II pointing contributing to the CTIO $Rz'$ data set
covers a field of 0.36 deg$^2$. For clusters located within a given
CTIO pointing, we employ the area of each pointing not associated with
the particular cluster detection (typically 0.33 deg$^2$) as the
control field. The cluster area itself is a circle of radius identical
to that applied to determine the colour--magnitude relation and the
colour histogram (typically 2 arcminutes radius) as detailed in
Section 3.1.

\begin{figure*}
\psfig{figure=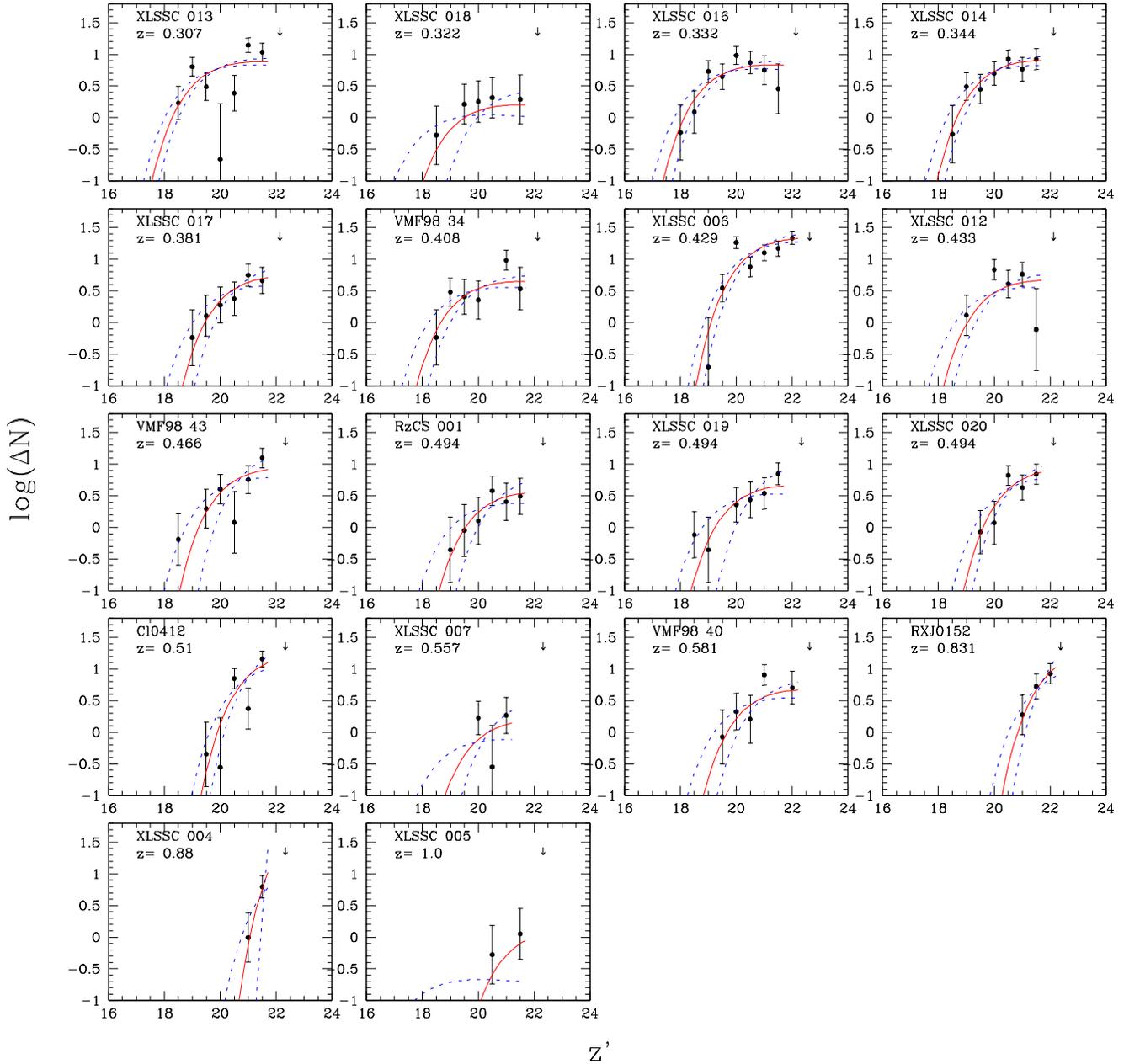,width=18truecm}
\caption[h]{Cluster galaxy luminosity functions computed in the $z'$
band. The ordinate is the logarithm of the number of galaxies per
bin. The arrow marks the magnitude completeness limit. The red (solid)
curve marks the best fit Schechter function, while the blue (dashed)
curve marks the best fit $\pm 1 \sigma$.}
\end{figure*}

\begin{figure*}
\psfig{figure=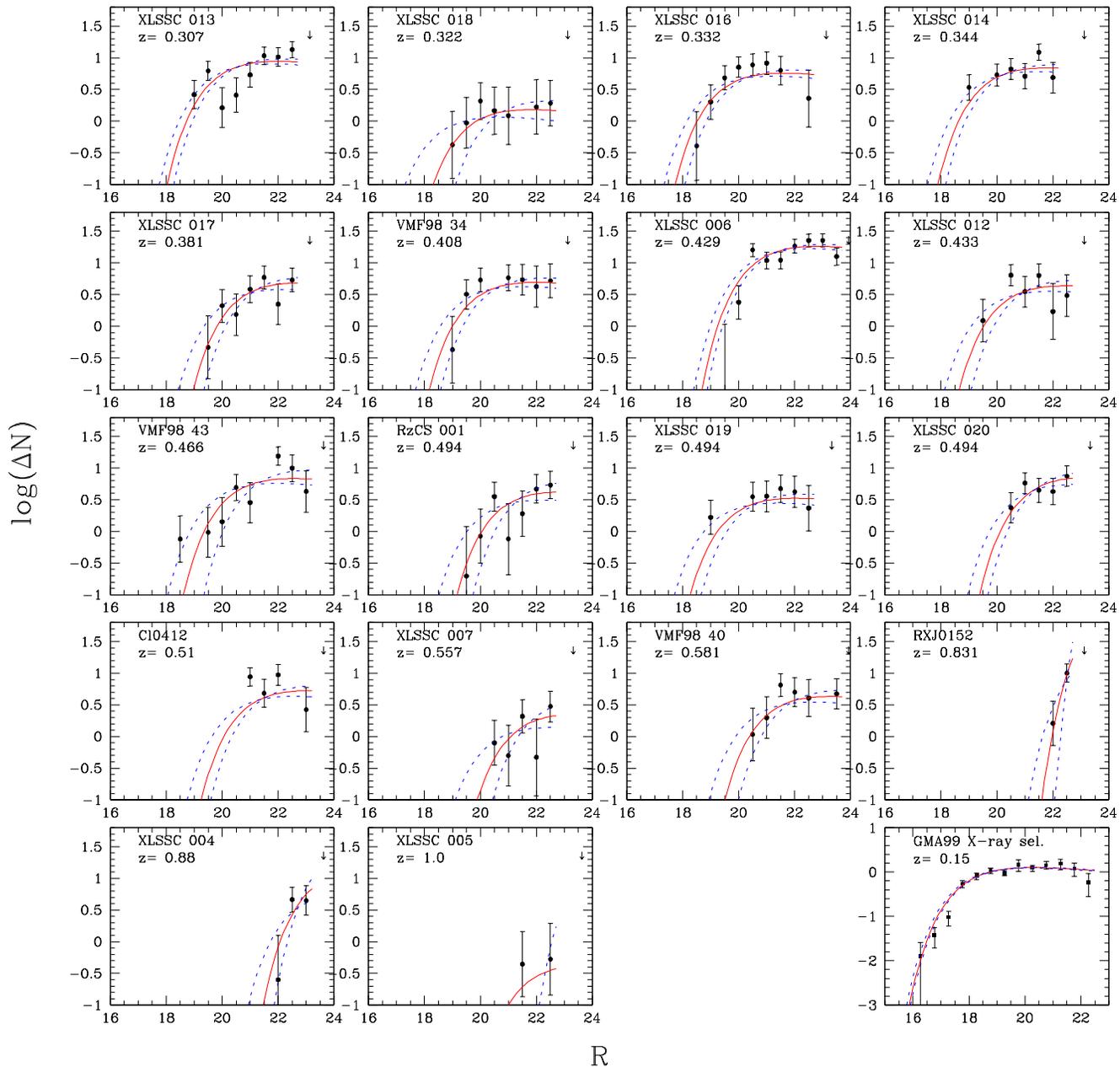,width=18truecm}
\caption[h]{Cluster galaxy luminosity functions computed in the $R$
band. Axis, symbols and curves as in previous figure.  The
bottom--right panel shows the low redshift R band LF of a composite
sample of 21 X--ray selected clusters.}
\end{figure*}

\begin{figure*}
\psfig{figure=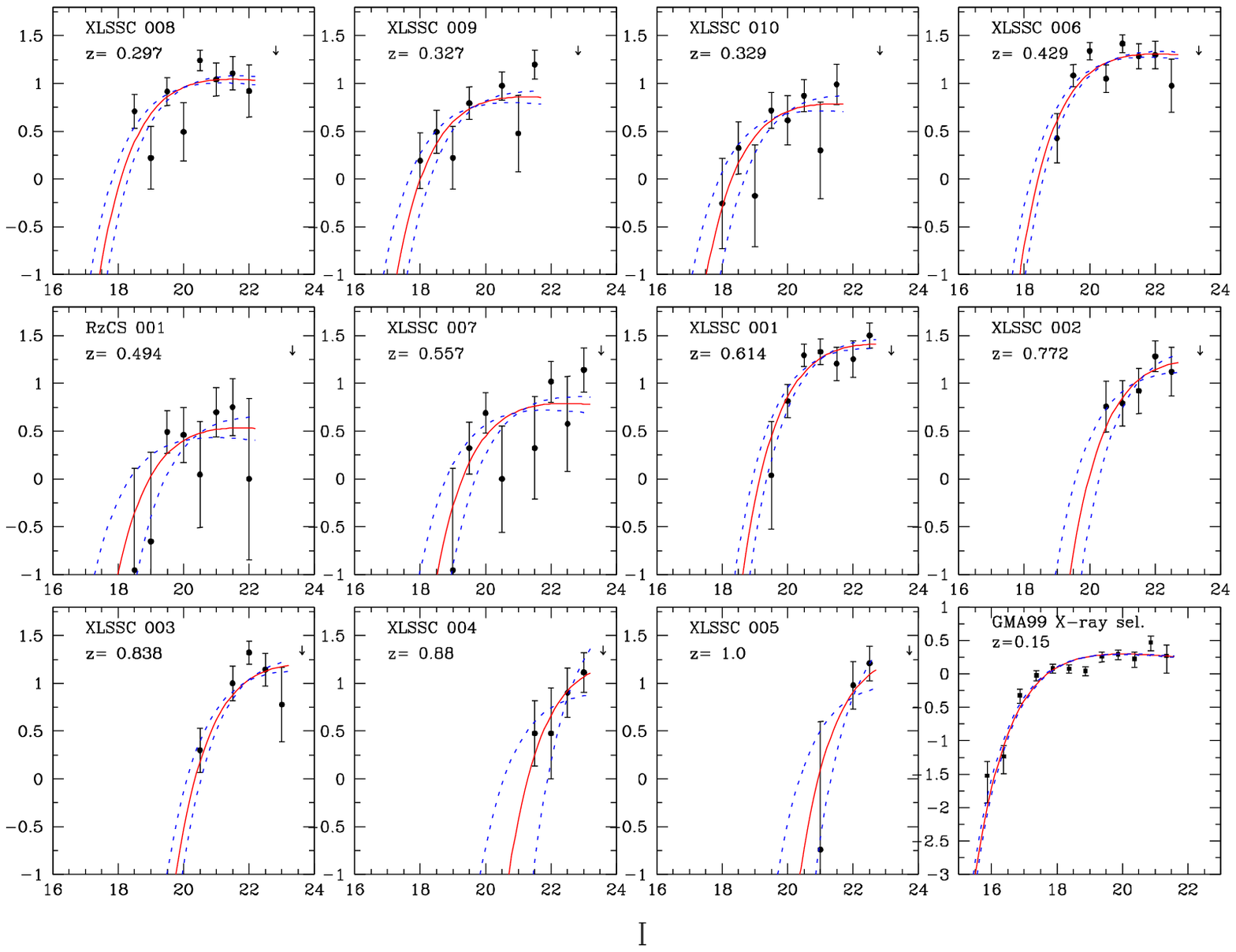,width=18truecm}
\caption[h]{Cluster galaxy luminosity functions computed in the $I$
band.  Axis, symbols, curves, and low redshift sample (bottom--right
panel) as in previous figure.  }
\end{figure*}

For clusters observed with VLT/FORS2, a circular aperture of 2
arcminute radius is applied to study the LF. Three exceptions are
clusters XLSSC 002, XLSSC 003 and XLSSC 004, for which the adopted
radius is 1.5 arcminutes.  The choice of the control field for
clusters studied employing the VLT data is more complex than the CTIO
case. At redshifts $z>0.6$ the apparent angular extent of the cluster
is small and the projected cluster galaxy surface density expected at
2.5 arcminutes from the cluster centre is assumed to be negligible.
Therefore, for such distant clusters, the area contained within a
circular annulus of radius 2.5 to 3 arcminutes is used as a control
field.  At redshifts $z<0.6$ the apparent angular extent of typical
clusters contaminate most of the FORS2 field of view, making the
definition of any cluster--free control area within the field
impossible.  For clusters located at redshifts $z<0.6$, we employ the
FORS2 field of cluster XLSSC 005 (located at $z=1$) as the control
field.  The decision is motivated by the fact that the brightest
member galaxy of XLSSC 005 is fainter than the faintest luminosity bin
of clusters located at redshifts $z\sim0.3$, i.e. XLSSC 005 is too
distant to affect the background count level at the magnitudes sampled
in lower redshift clusters.  For clusters located at redshifts
$z\sim0.5$, only the faintest magnitude bin is affected by any
contribution from the $z=1$ cluster.  As an additional test, we also
use a FORS2 image taken as part of the same data set of the field of a
redshift $z\sim1.3$ cluster candidate to compute the background count
level.  In this case there is little doubt that the distant cluster LF
will not bias the computation of the background level for $z\sim0.5$
clusters.  Comparison of the $z<0.6$ cluster LFs computed using the
above two background count models indicates no significant
differences.

Clusters and control fields are at most located within 1 degree of
each other (for VLT data), and the majority display separations of the
order of a few arcminutes.  Paolillo et al. (2001) demonstrated that
the only effect of selecting a background region too close to the
cluster is to reduce the S/N ratio of the resulting cluster LF,
without otherwise altering the resulting $m^*$ parameter.

The LF computation employs a parametric form described by a Schechter
(1976) function
\begin{equation}
\phi(m)=\phi^*10^{0.4(\alpha+1)(m^*-m)}e^{-10^{0.4(m^*-m)}},
\end{equation}
where $m^*$ and $\alpha$ are the characteristic magnitude and the
slope of the LF at faint magnitude respectively. The LF normalization,
$\phi^*$, is not constrained in the fitting procedure.  The value of
the faint--end slope of the Schechter function fitted to the data is
fixed at $\alpha=-0.87$ (the value derived at low redshift by
GMA99). The slope is fixed as it is relatively unconstrained at high
redshift and, as $m^*$ and $\alpha$ are highly correlated, performing
an unconstrained fit with respect to $\alpha$ will seriously undermine
any constraints placed upon $m^*$. This approach introduces the
drawback that any differential luminosity evolution occurring between
bright and faint galaxies within a given cluster will be ignored
(Andreon 2004). We further specify that the integral of the model LF
over the observed magnitude range be equal to the observed number of
galaxies, leaving $m^*$ as the only free parameter.  We also take into
account the finite width of magnitude bins by convolving the model
function with a rectangular window function of 0.5 mag width.

The $\chi^2$ statistic is computed as
\begin{equation}
{
\chi^2 = \sum_{bins} (Model-Observations)^2 / error^2,
}
\end{equation}
where $error^2$ is the quadrature sum of the Poisson uncertainty of
the $Model$ and background contributions. We adopt a theoretical
definition of $\chi^2$ in order to include the information in
magnitude bins containing no galaxies.  Errors on $m^*$ corresponding
to 68\% confidence levels are computed from $\chi^2=\chi^2_{best}+1$
(Avni 1976; Press, Flannery, \& Teukolsky 1986).

The LFs presented in this paper are computed employing a specific
metric aperture to determine galaxy magnitudes (Section 2).  The
application of a metric aperture permits a consistent comparison to be
made between the LFs computed at different redshifts.
However, the LFs presented in this paper are not comparable to 
LFs computed using different photometric apertures, e.g. isophotal or
pseudo--total photometric measures (as discussed in Section 2).

\subsubsection{$R$-- and $z'$--band LFs}

Computation of the LF for high--redshift clusters is dependent upon
additional cluster member selection techniques (i.e. colour selection
of the red sequence) as the cluster contribution is sometime
numerically small compared to the galaxy ``background''.  Therefore,
cluster $R$-- and $z'$--band LFs are computed applying a selection in
colour space, i.e. selecting photometric cluster members within a
specific colour interval.  The applied interval (with exceptions
mentioned below) was 0.65 mag in $R-z'$ delimited on the red side by
the colour of the red sequence plus 0.05 mag -- to account for the
broadening of the colour--magnitude relation by photometric
uncertainties. The applied colour interval is indicated in Figure 3 by
vertical arrows and in Figure 2 by horizontal lines (note that the red
selection limit is redder than the red sequence marked in each figure
by 0.05 mag -- as outlined above). Rejecting galaxies redder than the
red envelope cut--off removes galaxies that are too red to be
plausibly located at the cluster redshift. No overdensity of galaxies
redder than the applied colour limit is observed within any of the
cluster fields (see Figure 3), except where a background cluster is
located along the line of sight, e.g. XLSSC 017.  Fukugita et
al. (1995) indicate that the applied blue selection limit excludes
galaxies bluer than ``Irregular--type'' at redshifts $z>0.5$ and
excludes no galaxy types at lower redshift.

The blue colour limit includes the colour range covered by the
observed galaxy overdensity and no statistically significant galaxy
overdensity is detected blueward of the blue colour limit for the
majority of the clusters presented in this sample. Two exceptions are
XLSSC 020, where XLSSC 017 is located in the foreground, and XLSSC
007, which is not detected as a significant optical overdensity and
for which the colour selection interval is reduced to exclude
spectroscopically confirmed interloping galaxies.

The spectroscopic galaxy sample permits an independent check that the
applied colour selection criteria do not introduce a significant level
of incompleteness into the cluster galaxy sample.  We found that the
applied colour selection interval excludes on average only 10\% of
spectroscopic cluster members. This figure drops to 5\% if the cluster
XLSSC 007, which was assigned a smaller colour interval due an unusual
background, is excluded. Therefore, the applied colour selection
criteria do not exclude an important population of potential cluster
members and the procedure does not bias the computation of the $R$--
and $z'$--band LF. As a final test, the $R$-- and $z'$--band LFs for
each cluster computed without applying colour selection are compared
to the LFs generated by the colour selected cluster galaxy sample.

Figures 5 and 6 display the $z'$-- and $R$--band LFs for each
cluster. The LFs are computed applying the above colour selection
criteria. If colour criteria are not applied, statistically identical
results are obtained for all but the two highest redshift clusters in
the sample, plus clusters XLSSC 007 and XLSSC 020 -- not unexpected
given the above discussion.  The limiting magnitude sampled in each
cluster LF is $m<m^*+3$ in $z'$ and $m<m^*+4$ in $R$ for low redshift
clusters decreasing to $m<m^*$ for high redshift clusters.  The plots
indicate the best fitting Schechter model (solid red line) and the
models corresponding to a $\pm 1 \sigma$ variation in $m^*$ (dashed
blue line). In each case a Schechter function provides an acceptable
representation of the LF data at the 99 \% confidence level or
greater.  All $R$-- and $z'$--band LFs display a shift to fainter
apparent magnitudes with increasing redshift (as expected).  Certain
clusters require additional comments:

\begin{enumerate}

\item{XLSSC 018 is an intrinsically poor cluster. We count only 8
galaxies brighter than the $z'$--band limiting magnitude within 2
arcminutes of the cluster centre. Several spectroscopically confirmed
members are located outside this radius (a total of 12 galaxies have
been confirmed spectroscopically as being members of this cluster).}

\item{XLSSC 006 is one of the optically richest clusters in the
sample, with more than 77 photometric members within the 2 arcminute
radius aperture and brighter than the $z'$--band magnitude limit. By
counting the galaxies within the interval $m_3$ to $m_3+2$, and
accounting approximately for the population outwith the nominal
cluster radius, the cluster displays an Abell (1957) richness class of
0.  The other clusters in the sample at similar or lower redshift
display lower optical richness values. The aberrant $z'$ LF data point
at $z'=20.0$ is deviant from the Schechter LF by several sigma, and it
is therefore it is rejected from the $\chi^2$ fitting. No other points
are similarly aberrant in any of the presented clusters, and no other
LF data point is rejected.}

\end{enumerate}

The lower--right panel of Figure 6 displays the LF computed for 21
X--ray selected clusters at redshifts $z<0.25$ (GMA99). The
transformation from the $r'$--band employed by GMA99 and the $R$--band
employed for the current sample is performed by applying $R-r'$ values
following Fukugita et al. (1995). One third of the GMA99 sample is
X--ray selected, while half of the remaining sample is composed of
clusters subsequently detected in the Rosat All Sky
Survey\footnote{This is demonstrated by comparing the GMA99 and
Cruddace et al. (2002) cluster lists.}.  The $R$--band LF derived
using the GMA99 sample of 65 clusters is identical within the errors
to the subset of 21 X--ray selected GMA99 clusters.

At the median redshift of the sample, $\langle z \rangle = 0.47$, the
$R$--band samples rest--frame wavelengths $\lambda \sim 4000$ \ \AA,
whereas the $z'$ band samples the rest--frame $V$--band.  At such
redshifts, the relative magnitude change arising from recent or
continuing star formation will be greater than at rest--frame
wavelengths $1.4 \la \lambda (\mu m) \la 1.6$ sampled by NIR ($K$--band)
passbands.

\begin{figure*} 
\psfig{figure=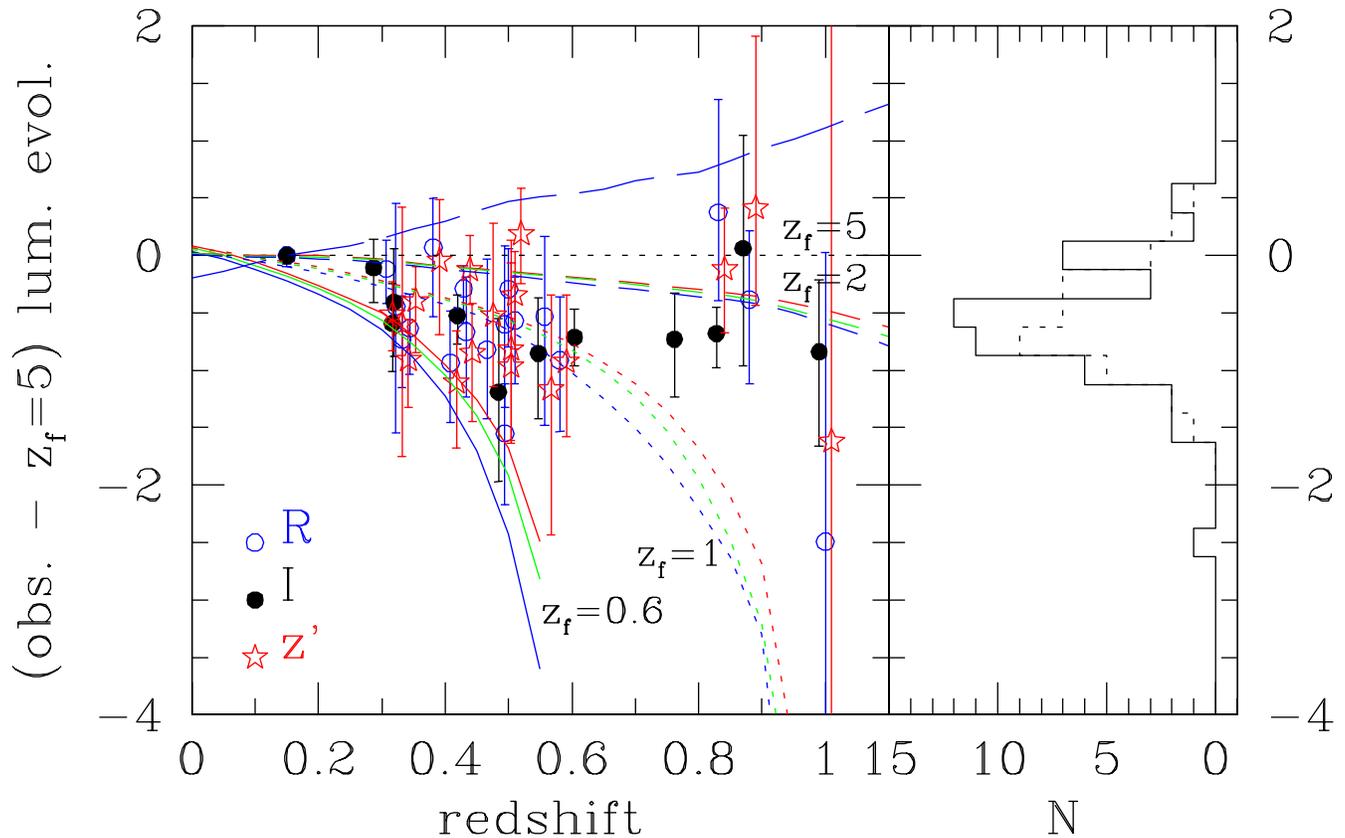,width=18truecm}
\caption[h]{{\it Left panel:} Characteristic LF magnitude, $m^*$,
evolution as a function of redshift having removed the contribution
from passive ($z_f=5$) stellar evolution, and from luminosity distance.
The labelled curves are the
predictions for different formation redshifts. Points and curves of
the same colour refer to the same filter, as indicated within the
figure. The long--dashed curve is the $R$--band expectation neglecting 
stellar evolution. Points are slightly offset in redshift in order to limit
crowding. The R and I calibrating points at $z=0.15$ (mean of 21 clusters)
fall one on the top of the other and hence are not easily to see separately.
{\it Right panel:} frequency distribution of the
points in the left panel (solid histogram) and of the corresponding
values derived without any colour selection and excluding problematic
clusters.} \end{figure*}

\subsubsection{$I$--band LF}

Figure 7 displays the $I$--band LFs computed for the cluster sample.
In contrast to the $R$-- and $z'$--band LFs discussed in previous
sections, no colour selection was applied to generate the $I$--band LF
sample, either because no $Rz'$ data was available, or, in the case of
clusters common to both data sets, because of large differences in the
field size between filters.

The lower--right panel displays the composite LF computed for 21
X--ray selected clusters at $z<0.25$ (GMA99), converted from $i$ to
$I$ using Fukugita et al. (1995), whereas the other panels
show the LFs computed for individual clusters presented in this paper.
The curves show the best fitting Schechter model (solid red line) and
models corresponding to a $\pm 1 \sigma$ variation in the $m^*$
(dashed blue line).  The Schechter function provides an acceptable fit
to the LF data points at the 99 \% confidence level.  The limiting
magnitude sampled in each cluster LF runs from $m<m^*+3$ at low
redshift to $m<m^*+1$ at the highest redshift in the sample.

\subsection{Global luminosity evolution}

Values of $m^*$ and corresponding uncertainties computed for cluster LFs
in each of the three filters considered are 
available in electronic form 
\footnote{at the URL http://www.brera.mi.astro.it/$\sim$andreon/XIDindex.html}. LF parameters for the composite LF of the GMA99
redshift $z=0.15$ sample are also tabulated.

Figure 8 displays the redshift dependence of 47 $m^*$ values generated
by LF computations for clusters in our sample, plus two $z=0.15$
reference points (from GMA99), each one being the average of 21 X--ray
selected clusters.  To highlight the possible effects of active
luminosity evolution upon $m^*$, the distance modulus and passive
luminosity evolution terms were removed assuming a passively evolving
stellar population formed at $z_f=5$ using the model of Kodama \&
Arimoto (1997) -- the model previously employed to compute the colour
of the red sequence.  The model predictions are normalized to the
observed $m^*$ at $z=0.15$ in $R$ and $I$.  A local determination of
the $z'$--band cluster LF is not currently available. We therefore
adopt the computed $I^*$ value for the local cluster sample and
applied $I-z'=0.2$ mag (Fukugita et al. 1995).  The measured $R$--,
$z'$-- and $I$--band $m^*$ values, once passive evolution has been
accounted for, do not differ systematically from each other -- further
confirmation that the colour selection applied to compute the $R$--
and $z'$--band LFs, but not employed for the $I$--band, has little
impact on the computed value of $m^*$.

The $m^*$ data points are systematically brighter than a model based
upon an old, passively evolving stellar population (the horizontal
dashed line in Figure 8). A non--evolving model is also strongly ruled
out.  The clusters presented in Figure 8 require the occurrence of a
secondary star formation episode at lower redshift than the $z_f=5$
in order to generate $m^*$ values brighter than the
passive evolution model. The same conclusion can be drawn by
considering the right panel of Figure 8 (solid line) which displays
the histogramme of $m^*$ values summed over redshift. The resulting
$m^*$ distribution is approximately 1 magnitude wide and is offset
from zero.  The dotted histogramme in the same panel displays the
$m^*$ histogramme computed for cluster LFs generated without colour
selection and by rejecting problematic clusters (i.e. clusters blended
along the line of sight and those displaying $m^*$ errors larger than
0.6 mag), demonstrating again that the applied colour selection does
not introduce a significant bias.

Paolillo et al. (2001) and Andreon (2004) 
demonstrate that the LF shape computed for
galaxies drawn from extended cluster regions are statistically equal
to the corresponding LF shape computed for galaxies drawn only from
the central cluster regions. Therefore, the specific choice of cluster
aperture radius employed to generate the cluster LF sample should not
introduce a significant bias into the resulting LF computation. 
GMA99 and Paolillo et al. (2001), both show that the slope of the 
composite LF of many clusters, in the magnitude
range sample in these works and in the present paper, does not depend
on wavelength from $g$, or $B_j$, to $i$. Christlein, McIntosh \& Zabludoff 
(2004) shows the similarity of the slopes in the $U$ and $R$ bands
of the composite LF of three clusters. Therefore, our results should
not be biased by a wavelength--dependent slope. A
remaining concern may be that the reference $m^*$ value at low
redshift (i.e. the GMA99 data) could be inappropriate for the current
sample of (typically) low richness clusters, as the GMA99 sample
includes some optically richer clusters. The right panel of Figure 8
indicates that, in order to remove the supposed luminosity evolution,
$m^*$ should depend on optical richness as much as 1 mag when
considered across the range of richness values displayed by clusters
in the current sample. Within the large GMA99 sample there is a
$0.01\pm0.15$ mag difference in $m^*$ between rich and poor
clusters. This indicates that, while the $m^*$ normalization as a
function of richness may play some role, it is unlikely to account for
the apparent evolution in $m^*$ with redshift.

The additional curves in Figure 8 indicate the expected $m^*$
evolution for stellar populations formed at successively lower
redshifts. In order to account for the bright $m^*$ values observed at
redshifts $z \sim 0.3$ a formation redshift as low as $z_f=0.6$ would
be required, although, by adopting such a low formation redshift, the
predicted $m^*$ value at a redshift $z=0.15$ would be 0.2 magnitudes
brighter than that reported by GMA99. The LF data points are not well
described by any formation model based upon a single episode of star
formation and at least two important star formation events are
required.  The last (in cosmic time) star formation event should
brighten average $m^*$ values by up to 1 magnitude (right panel of
Figure 8). Such secondary star formation activity may be related to
the Butcher--Oemler effect (Butcher \& Oemler 1984), although the
evidence for the latter is not compelling (Andreon \& Ettori 1999;
Andreon, Lobo \& Iovino 2004). A Butcher--Oemler analysis of the
present sample of clusters is presently in progress.

A similar study by de Propris et al. (1999), performed for optically
rich clusters in the $K$--band, reported a redshift evolution of $m^*$
consistent with the prediction of a passively evolving stellar
population. In contrast, the current data set appears to indicate that
a secondary star formation episode is required. However, for clusters
considered to redshifts $z<1$, the $K$--band samples the rest--frame
contribution of old stars when weighted by luminosity. The secondary
star formation episodes supported by the current data set compiled
with red optical (rest--frame blue) passbands would not result in a
strong signal in the $K$--band cluster LF evolution. We note also that
a large dispersion is present in the $m^*$ values presented by de
Propris et al. (1999), and that several data points are brighter by 2
$\sigma$ than a prediction considering a passively evolving stellar
population formed at a redshift $z_f=3$.  There is therefore no
contradiction between the de Propris et al. (1999) LF evolution and
the results on luminosity evolution in clusters presented in the
current paper.

The sample of cluster LF $m^*$ values presented in this study are not
consistent with the prediction of a single evolutionary model, in
qualitative agreement with 
Dahl{\' e}n, Fransson, {\" O}stlin, \& N{\" a}slund (2004).
Therefore, not all clusters share the same evolutionary
history, for example because composed of different proportions
of passive and active evolving galaxies,
leading to a distribution of LF properties.  Under such
circumstances the computation of a composite LF for a sample of
cluster would be of questionable merit. Compilation of a large sample
of clusters to redshifts up to $z\sim1$ according to well--defined criteria
will permit a detailed investigation of the galaxy cluster LF and
sources of dispersion therein. This aim represents one of the
scientific goals of the continuing XMM--LSS survey.

\subsection{Comparison with previous works}

\subsubsection{Barrientos \& Lilly (2003)}

Barrientos \& Lilly (2003, hereafter BL03) present a study of the
luminosity and colour properties of 8 galaxy clusters located within
the redshift interval $0.40<z<0.48$. The authors present $I$--band
cluster LFs and a composite $V-I$ colour magnitude diagram. One of the
central claims of their paper is that the characteristic magnitude
$m^*$ of the cluster LF evolves passively -- a conclusion apparently
at variance with that presented in this work. Specifically, comparison
of the BL03 and our LF analysis indicates an offset in the
characteristic magnitudes evolution at the level of 0.5 magnitudes for
the two cluster samples drawn from an overlapping redshift
interval. This section addresses the causes of this discrepancy.
BL03 
reported a best fitting characteristic magnitude of $I^*\sim19.3$ for
red cluster galaxies, generally in good agreement with values computed
in the current paper of $I^*=19.2$ at $z=0.33$ to $I^*=19.6$ at
$z=0.42-0.49$. However, one should note that the computation
presented in the current paper includes galaxies of all colour, and
that the two samples employ marginally different faint--end slopes.

In order to constrain the brightness evolution of the LF, BL03 convert
apparent $I^*$ values to rest--frame, absolute $M_V$ values assuming a
cosmological model described by the parameters $q_0=0.5$ or $0.1$ and
$\Omega_{\Lambda}=0$.  When compared to the cosmological model considered in
this work (and supported by current observations), the BL03 assumed
model results in galaxies appearing fainter by $0.37$ or $0.18$
magnitudes respectively, than in our cosmological model.

In addition, to convert from observed $I$ to rest--frame $V$
magnitudes, BL03 employ the $V-I$ colour of a present day elliptical
galaxy.  Figure 4 indicates that the reddest galaxies at $z=0.4$ are
0.2 magnitudes bluer in $R-z'$ compared to a present day elliptical
galaxy.  Kodama et al. (1998) show that the same holds true in the
$V-I$ colour. Hence, the $V-I$ colour assumed by BL03 is
0.2 mag too red.

Therefore, in computing absolute $M_V$ values from apparent $I$--band
magnitudes observed for elliptical galaxies at typical redshifts
$z\sim 0.45$, BL03 introduce a total systematic offset of $0.2 +
(0.37,0.18)$ magnitudes in the sense that their final $M_V$ values for
galaxies at $z\sim0.4$ are fainter than values computed employing
observed colours at $z\sim0.45$ and a $\Lambda$ dominated universe.

In order to constrain the amplitude of luminosity evolution from a
redshift $z\sim0.45$ to $z=0$, BL03 employ a sample of low--redshift
clusters compiled by Lopez-Cruz (2001).  Formation of a consistent
comparison is hindered by the fact the LF parameters for the
Lopez-Cruz (2001) sample are computed independent of cluster galaxy
colour (recalling that the BL03 sample is restricted to
colour--selected early--type galaxies) in addition to the requirement
to apply further corrections to account for the different photometric
filter response functions. Most importantly, the median redshift of
the Lopez-Cruz (2001) cluster sample is $z\sim0.1$. However, the
``local'' ($z=0$) LF parameters derived from Lopez-Cruz (2001) neglect
the effects of passive stellar evolution from $z=0.1$ to $z=0$.  In
simple terms, Lopez-Cruz (2001) measures $m^*$ at $z=0.1$ (a general
discussion can be found in Andreon 2004).  The passive evolution
expectation should be, therefore, computed from $z_{low}=0.1$ to
$z_{high}\sim0.45$, whereas BL03 take $z_{low}=0$. Because of
overestimation of the redshift baseline, BL03 introduce a further
magnitude offset of 0.1 magnitudes in the sense that the apparent
evolution thus computed is underestimated by 0.1 magnitudes, as
directly measured by Blanton et al. (2003) and Andreon (2004), and in
agreement with passive evolution models (e.g. Bruzual \& Charlot
1993).

We therefore conclude that BL03 underestimate the luminosity evolution
within their cluster sample by 0.67 to 0.48 magnitudes (with the two
values generated by the two cosmological models assumed in BL03),
exactly the discrepancy between the $m^*$ evolution of the BL03 sample
and the current work. Application of these corrections to the BL03
data indicates a $R$--band luminosity evolution amplitude of 1.4
magnitudes (with errors greater than 0.5 magnitudes) compared to an
expected passive evolution amplitude of approximately 0.5
magnitudes. We conclude that the BL03 data do support an additional
luminosity evolution term in excess of that expected on the basis of
passive evolution alone. Based upon these arguments we claim that the
`active' luminosity evolution claimed in the current paper is observed
in the BL03 sample.

\subsubsection{Nelson et al. (2001).}

Nelson et al. (2001; hereafter N01) compute $m^*$ values for the
$I$--band LF distributions of 12 clusters with spectroscopic redshifts
in the range $0.35<z<0.65$ (see their Table 2). Comparison of the
cluster luminosity distributions presented in N01 with those presented
here indicate that observations of the former are shallower or comparable
to our ones. When the N01 sample is augmented by literature data,
Nelson et al. report a variation of 1.65 magnitudes in $m^*_I$ in the range
$0.35<z<0.85$.  The passive evolution model prediction in the same
redshift range is 2.65 magnitudes, i.e. the observed characteristic
magnitude is about one magnitude brighter compared to the passive
expectation -- in excellent agreement with our findings.
However, N01 report a different conclusion, stating that $m^*$
is evolving passively according to a stellar population formed at a
redshift $z_f=1.7$ (right--hand panel of their Figure 11), because N01 employ
an older cosmological model. Their data,
in the present cosmological model, are well described by the evolution
of a stellar population forming at $z_f\sim 0.7$, in good agreement
with our claim of a secondary (i.e. below $z<1$) star formation
activity pointed out by our data. We note that a direct comparison of
LF values is not possible as a result of the
particular LF fitting procedure adopted by N01.

\section{Discussion and conclusions}

We have studied a sample of 24 clusters located at redshifts
$0.297<z<1$, of which 16 display redshifts $z>0.4$ and 6 have
$z>0.6$. 
The majority of the clusters are either X--ray selected or
detected, and we are therefore observing gravitationally bound
systems. Most of the cluster sample, particularly clusters 
at redshifts $z<0.6$, possess X--ray luminosities and optical richness
values typical of groups or low mass clusters.

All clusters in our sample, despite the primary X--ray selection and low
X--ray flux/optical richness displayed by the majority of the sample,
display a
statistical overdensity of galaxies of similar colour (Figure 3), that
make them detectable by an almost 3
dimensional search defined by sky position and colour. In fact,  
all clusters
with $R$ and $z'$ photometry, with the exception of XLSSC 007, are
colour--detected.  However, the present optical
identification of XLSSC 007 as the counterpart of the extended X--ray
source is uncertain as a result of the large distance between the
optical overdensity and X--ray centres.  Should the identification of
this X--ray source change to that of a $z>1$ cluster, then no X--ray
selected cluster presented in this paper is missed by the $R-z'$
technique in the $z<1$ regime. Most of the clusters are identified in
X--rays, largely independent of the optical luminosity of the member
galaxies.  Therefore, the colour detection is non--trivial. The majority
of the clusters are optically poor (Abell richness class 0 or
lower) consistent with the low computed X--ray luminosities. We have
therefore demonstrated that a colour plus spatial overdensity search
technique can effectively identify optically poor systems at
intermediate to high redshifts (at least those previously identified
in X--rays).

The emerging picture from the current study is the one of a typical
cluster composed of two or more distinct galaxy populations :
a relatively old population evolving passively (as measured from
the evolution of the color of the red sequence) together with a
younger population, ostensibly responsible for the apparent
brightening of the characteristic LF magnitudes. 

The reddest galaxies
within each cluster/group evolve in a manner consistent with a model
early--type galaxy formed between redshifts $2\la z_f \la 5$ (Figure
4). This observation is largely in agreement with previous studies. We
note, however, that previous studies employ cluster samples
dominated by optically rich systems often observed with heterogeneous
instruments. In contrast, the current study consists of an
exceptionally uniform cluster sample observed under largely uniform
conditions. Previous studies estimate similar values for $z_f$, though
assuming different cosmological models, evolutionary models, or
both. We note that the formation epoch of cluster galaxies estimated
for the current sample would correspond to a higher redshift
for the same assumptions adopted in literature.

The younger population is detected by studying the LF.
The LF of each cluster has been computed in $R-, I-$ and $z'$--bands
and is displayed in Figures 5 to 7. A Schechter function provides an
acceptable description of the LF shape over the magnitude range
extending from $m<m^*$ to $m<m^*+4$, with exact values depending on
redshift and filter. The distribution of LF $m^*$ values versus
redshift is systematically brighter than predictions based upon a
passively evolving stellar population formed at $2 \la z_f \la 5$
perhaps because our redshifts are sampling ``the time of rapid
cluster building" (Dressler 2004).  The $m^*$ values are, on average,
almost one magnitude brighter than the passive evolution prediction --
indicative of {\it active} luminosity evolution, or secondary
star--formation activity. The $RIz'$ passbands used in this study
sample galaxy emission at typical rest--frame wavelengths
corresponding to the $B$-- and $V$--bands. The resulting LFs are
therefore more sensitive to the effects of recent star formation than
cluster studies over similar redshift intervals employing NIR
passbands. The evolution of $m^*$ provides a measure of the evolution
of the {\it whole} galaxy population, as opposed to that derived from
the colours of the reddest galaxies that monitor the evolution of the
oldest cluster galaxies. Overall, the galaxy population is
actively evolving.

Therefore, we have detected two distinct galaxy populations, one passively
evolving and another one actively evolving.
The determination of the nature of this secondary activity (e.g. the
time scale and the relationship with the cluster properties and the
identification of the active evolving population) is within
the reach of the XMM--LSS project, since the $z<1.3$ redshift regime
will be ultimately sampled by several hundreds of X--ray selected
clusters with supporting multi--colour and spectroscopic observations.

\section*{Acknowledgements}
We thank T. Kodama for providing his model predictions, C. Mullis for
providing galaxy redshifts in advance of publication, L. Jones,
M. Paolillo for their comments to the earlier version of
this papers and A. Dressler and collaborators of the XMM--LSS project 
for stimulating conversations.  The referee is acknowledged for 
pointing out a literature paper initially forgot by us and
for comments that improved the paper presentation.
H.Q. thanks the FONDAP Centro de Astrofisica for
partial support and the award of a Guggenheim Foundation fellowship.
SA has received support from MURST-COFIN n. 2003020150-005.

\bsp

\label{lastpage}

\end{document}